\documentclass[letterpaper,12pt]{article}   
\usepackage{osajnl2} 
\usepackage{setspace}

\begin{document}

\title{Identification of competing ultrafast all-optical switching mechanisms in Si woodpile photonic crystals}

\author{Philip J. Harding$^{1, 2}$, Tijmen G. Euser$^{1, 3}$, and Willem L. Vos$^{1, 2}$}
\address{$^{1}$Center for Nanophotonics, FOM Institute for Atomic and Molecular Physics (AMOLF), Kruislaan 407, 1098 SJ Amsterdam, The
Netherlands}
\address{$^{2}$Complex Photonic Systems (COPS), MESA+ Institute for Nanotechnology, University of Twente, 7500 AE Enschede, The
Netherlands}
\address{$^{3}$Currently with Max-Planck Research Group (IOIP), University of Erlangen-Nuremberg, G\"{u}nther-Scharowsky-Str. 1/Bau 24, 91058
Erlangen, Germany}

\date{\today}

\begin{abstract}
We present a systematic study of ultrafast all-optical switching of Si photonic band gap woodpile crystals using broadband tunable nondegenerate
pump-probe spectroscopy. At pump-probe coincidence, we investigate the behavior the differential reflectivity at the blue edge of the stopband
for a wide range of pump- and probe frequencies. Both dispersive and absorptive features are observed from the probe spectra at coincidence. As
the pump frequency is tuned through half the electronic bandgap of Si, the magnitude of both these features increases. For the first time we
unambiguously identify this dispersive effect with the electronic Kerr effect in photonic crystals, and attribute the the absorptive features to
nondegenerate two photon absorption. The dispersive and absorptive nonlinear coefficients are extracted, and are found to agree well with
literature. Finally, we propose a nondegenerate figure of merit (NFOM), which defines the quality of switching for all nondegenerate optical
switching processes.
\end{abstract}

\ocis{050.5298, 190.3270, 190.4180, 320.7110, 320.2250}

\maketitle

\section{Introduction}
The interest to optically switch photonic structures has recently gathered momentum due to the inherent fastness of the process. While the
switching speed of conventional transistors is ultimately limited by heat dissipation, no such limitation exists for optical systems in the
absence of absorption. To date, the dispersion of photo-induced free carriers is widely exploited to induce ultrafast changes in the photonic
structure's optical properties. Then, propagation switching, optical modulation, trapping and releasing photons, frequency- and bandwidth
conversion, and even ultrafast switching of the density of states are possible \cite{Johnson:02, Bret:03, Almeida:04, Xu:05, Preble:07,
Harding:07}.

In the past, several groups have switched Si photonic crystals via free carrier excitation \cite{Leonard:02, BeckerAPL:05, Euser:07}. This
switching mechanism is popular because of its large attainable change in refractive index and its high possible repetition rate. Indeed,
currently many efforts are devoted to decreasing the timescales of switching in polycrystalline silicon (p-Si), achievable through implanting
additional recombination centers such as ions \cite{Chin:96, Foerst:07, Tanabe:07}. However, the possible recombination time is still limited to
several ps.

Electronic Kerr switching offers the potential to switch instantaneously, with a repetition rate limited only by the duration of the pump pulse.
The duration of the pump and probe pulses then controls the repetition rates, instead of the recombination times of the free carriers. Kerr
switching could potentially increase the repetition rate from GHz to THz. Unfortunately, the intensities required are typically much higher
compared to free carrier switching \cite{Dinu:03, Bristow:07, Lin:07}. Recently, several groups have claimed to have experimental evidence of
this Kerr nonlinearity in photonic crystals \cite{Hache:00, Mazurenko:03, Hastings:05, Mondia:05, KerrRetract:07}. The reasoning in all cases is
the instantaneous nature of the effect, resulting in a decrease in reflectivity or transmission at one probe frequency which coincides with the
cross-correlation of the pump and probe pulses. Here, we critically evaluate these optical processes at instantaneous (fs) timescales. At
pump-probe coincidence, we will show that two competing processes change the optical properties as witnessed from reflectivity measurements
performed over a large bandwidth. From these two processes, we propose a general nondegenerate figure of merit (NFOM), which has wide
implications for all-optical fs switching, ultimately extending to DOS switching and the dynamic control of spontaneous emission.

\section{Experimental setup and sample}
Our setup consists of a regeneratively amplified pulsed Ti:Sapph laser (Spectra Physics Hurricane) which drives two independently tunable
optical parametric amplifiers (OPAs, Topas). The OPAs have a continuously tunable output frequency between 0.44 and 2.4 eV, with nearly
bandwidth limited pulses (relative width of $1.33 \%$, bandwidth limited equivalent to $\tau_P = 110$ fs) with pulse durations of $\tau_P = 140
\pm 10$ fs (measured at $E_{\rm{Pump}} = 0.95$ eV) and a pulse energy of at least 20 $\mu$J. The probe beam is normally incident $\theta=
0^{\rm{0}}$ on the sample, and is focused to a Gaussian spot of 32 $\mu$m FWHM (at $E_{\rm{probe}} = 1.24$ eV) at a small angular divergence NA
= 0.02. The probe intensity $I_{\rm{Probe}}$ of $3 \pm 2$ GWcm$^{-2}$ was around 10 times lower than the pump intensity $I_{\rm{Pump}}$ to
ensure that nonlinear effects caused by the probe were negligible. The reflectivity was calibrated by referencing to a gold mirror. A versatile
measurement scheme was developed to subtract the pump background from the probe signal, and to compensate for possible pulse-to-pulse variations
in the output of our laser \cite{EuserWoodpile:08, EuserThesis:07}.

\begin{figure}
\begin{center}
\includegraphics[scale=0.25]{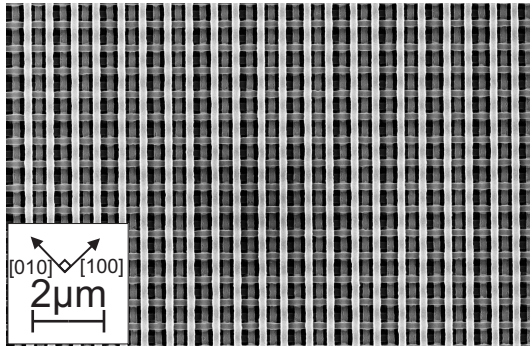}
\caption{\setstretch{2.0}High resolution scanning electron micrograph of the surface normal to [001] of a Si woodpile crystal at domain D4. The width and
thickness of each rod is $175 \pm 10$nm and $155 \pm 10$nm, respectively. SEMD4.eps}
 \label{fig:SEM}
\end{center}
\end{figure}

Figure \ref{fig:SEM} shows a high resolution scanning electron micrograph of the photonic woodpile structure made by Jim Fleming at Sandia
National Laboratories \cite{Fleming:99}. It consists of 5 layers of p-Si rods ($n'=3.45$) stacked orthogonally upon one another, the n-th layer
shifted by half an interrod distance with respect to the (n+2)th layer, with a total thickness of $L = 780$ nm. This structure gives rise to a
diamond lattice, for which a first-order band gap is predicted \cite{Ho:90}. The dimensions of the rods were chosen so as to aim the center of
the band gap around telecom frequencies ($E_{\rm{Tele}}=0.735$ eV). The electric field of the probe beam is polarized along the
[$\overline{1}$10] direction of the crystal, that is, perpendicular to the first row of rods. The last row of rods is supported by a 70 nm thick
SiN layer of refractive index $2$.

The sample has been characterized extensively elsewhere \cite{EuserThesis:07}. Over the sample, the mutual alignment of the rods differs. These
differences are spatially separated, and allow the sample to be divided into 16 domains, which we denote by A1 through D4 in analogy to a
chessboard. Here, we perform measurements on both A1 and D4, and are thus able to validate our results for different sample conditions. The
symmetry of D4 is face centered orthorhombic, while A1 is body centered orthorhombic.

\section{Linear reflectivity}
\begin{figure}[!h]
\begin{center}
\includegraphics[scale=1]{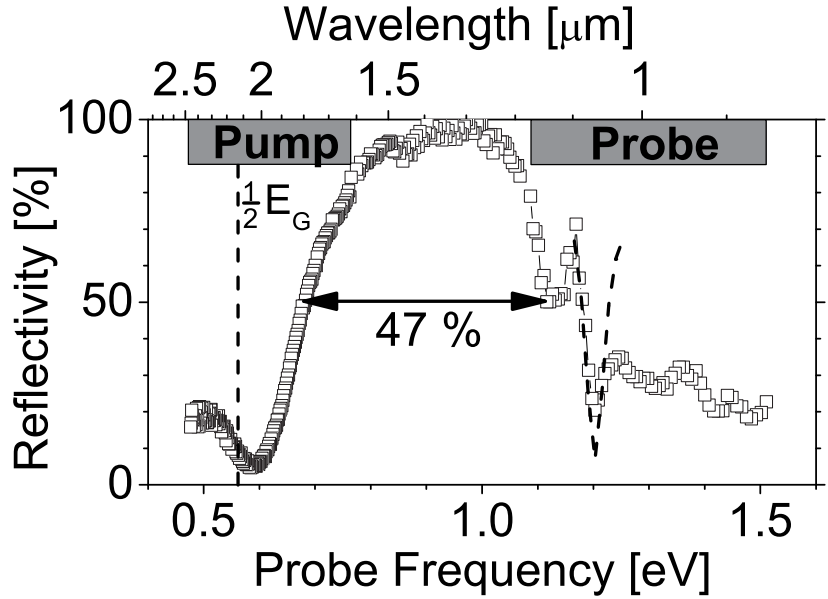}
\caption{\setstretch{2.0}Linear reflectivity spectrum of the woodpile photonic crystal measured normal to the [001] direction at a sample domain D4 shown in
figure \ref{fig:SEM}. The E-field is perpendicular to the first row of rods. A stopgap near 0.9 eV gives rise to a high maximum reflectivity of
$95 \pm 2 \%$ and has a broad relative width of 47$ \%$, indicating a high photonic strength. At high frequencies $> 1.2$ eV, the spectral
features are attributed to Fabry-P\'{e}rot-type fringes. The pump frequencies (shaded box) were tuned at the red edge through half the
electronic band gap $E_{\rm{G}}=1.12$ eV of silicon (vertical dashed line), and the probe frequencies at the blue edge of the stopband. The
dashed curve is a calculation with the Scalar Wave Approximation in the region of interest. ReflExtAllD4.eps}
 \label{fig:Refl}
 \end{center}
 \end{figure}

Figure \ref{fig:Refl} shows a linear reflectivity spectrum of the woodpile photonic crystal. The high peak centered at $E_0 = 0.9$ eV
corresponds to the $\Gamma-X$ stop gap in the band structure, which is part of the 3D photonic band gap of Si woodpile photonic crystals
\cite{EuserWoodpile:08, Gralak:03, Ho:94}. The maximum reflectivity of $95 \pm 2 \%$ and the broad width of $\Delta E=0.46$ eV FWHM (full width
at half maximum) confirm the strong interaction of the crystal with light. The interaction strength $S$ is characterized by the relative gap
width $S = \Delta E/E_0$ \cite{VosKreta:01}. We note that the large photonic strength of $S = 47 \%$ is the largest measured for any photonic
crystal so far.

We can associate a length scale with the photonic strength. The Bragg length $L_B$ is the length at which the incoming intensity has dropped to
$1/e$ of its initial value, and is given by \cite{KoenderinkThesis:03}
\begin{equation}
L_B = \frac{\lambda}{\pi S},
\end{equation}
where $\lambda$ is the wavelength in the structure of the stopgap. We deduce a Bragg length of $L_B = 840$ nm, corresponding to 1.06 unit cells.
In contrast, other 3D structures with a lower photonic strength have a much higher $L_B$: For example, TiO$_2$ inverse opals have $L_B \approx
3.3~\mu$m, polystyrene opals have $L_B \approx 5.1~\mu$m and SiO$_2$ opals $L_B \approx 6~\mu$m \cite{KoenderinkThesis:03}.

The high quality of the sample also causes the steep edges of the peak, which we will in first instance probe at the blue edge of the stopband.
At these frequencies, all induced changes in optical properties of the Si backbone will cause a large change in reflectivity, important for
switching applications. The probe frequencies (indicated by the shaded box) span both the blue edge of the stopband as well as the fringes.
These have been tentatively assigned to Fabry-P\'{e}rot-type interferences \cite{deDood:03}, but the interpretation of the reflectivity spectrum
of woodpiles is still an open question \cite{EuserThesis:07}.

\section{Switched reflectivity vs. time at one probe frequency}
\begin{figure}
\begin{center}
\includegraphics[scale=1]{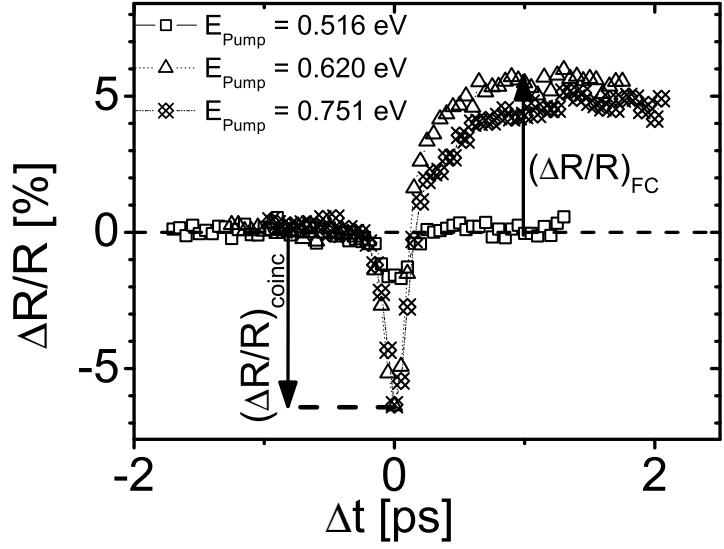}
\caption{\setstretch{2.0}Differential reflectivity $\Delta R/R$ versus probe delay $\Delta t$ taken at different pump frequencies $E_{\rm{Pump}}$ and at probe
frequency $E_{\rm{Probe}} = 1.13$ eV, sample domain A1. At $\Delta t = 0$ ps, the pump and probe are coincident, and the differential
reflectivity $(\Delta R/R)_{\rm{coinc}}$ decreases. At $\Delta t=1$ ps, the differential reflectivity $(\Delta R/R)_{\rm{FC}}$ has increased due
to the dispersion of the free carriers (FC). The peak pump intensity varies between $I_{\rm{Pump}} = 10 \pm 1$ GWcm$^{-2} (E_{\rm{Pump}} =
0.516$ eV) and $I_{\rm{Pump}} = 25 \pm 2$ GWcm$^{-2} (E_{\rm{Pump}} = 0.75$ eV), and the probe intensity was $I_{\rm{Probe}} = 3\pm 2$
GWcm$^{-2}$. DiffdRRvst.eps}
 \label{fig:timetraces}
 \end{center}
\end{figure}

The pump frequencies $E_{\rm{Pump}}$ were chosen as to tune through half the electronic band gap of silicon, where the gap is $E_{\rm{G}}=1.12$
eV. Both the Kerr coefficient $n_2$ and the degenerate two-photon absorption coefficient $\beta_{11}$ have recently been shown to vary strongly
in the vicinity of $\frac{1}{2}E_G$ \cite{Bristow:07, Lin:07, Euser:05, Reintjes:73, Tsang:02, Dinu:03}. On domain A1 we measured differential
reflectivity $\Delta R/R$ vs. probe delay $\Delta t$ as a function of $E_{\rm{Pump}}$ (figure \ref{fig:timetraces}). On this domain, the probe
frequency $E_{\rm{Probe}} = 1.13$ eV corresponds to the foot of the blue stopband edge. At positive probe delays $\Delta t=1$ ps and at pump
frequencies $E_{\rm{Pump}}>\frac{1}{2}E_{\rm{G}} = 0.56$ eV, we observe a large positive differential reflectivity $(\Delta R/R)_{\rm{FC}}$ due
to a blueshift of the photonic features. This shift is caused by a decrease of the refractive index of the p-Si backbone that agrees well with a
Drude description of the excited free carriers \cite{EuserWoodpile:08}. When $E_{\rm{Pump}}$ is reduced to below $\frac{1}{2}E_{\rm{G}}$, the
dispersion vanishes, since no free carriers are excited by a two-photon process.

At pump-probe coincidence, near $\Delta t=0$ ps, we observe a trough of magnitude $(\Delta R / R)_{\rm{coinc}}$. The width of this trough varies
between 240 fs ($E_{\rm{Pump}} = 0.52$ eV) and 160 fs ($E_{\rm{Pump}} = 0.62$ eV). These widths agree reasonably well with the expected
cross-correlation duration for two pulses ($\tau_P \times \sqrt{2} = 200$ fs). As $E_{\rm{Pump}}$ increases to above $\frac{1}{2}E_{\rm{G}}$,
the magnitude of the instantaneous feature $(\Delta R/R)_{\rm{coinc}}$ grows until it saturates at $E_{\rm{Pump}} = 0.62$ eV, just over
$\frac{1}{2}E_{\rm{G}}$. In all cases, instantaneous switching features are observed.

\section{Switched reflectivity vs. frequency}
\subsection{Reflectivity vs. frequency at coincidence}
\label{subsec:Rvsomegat0} 
\begin{figure}
\begin{center}
\includegraphics[scale=0.9]{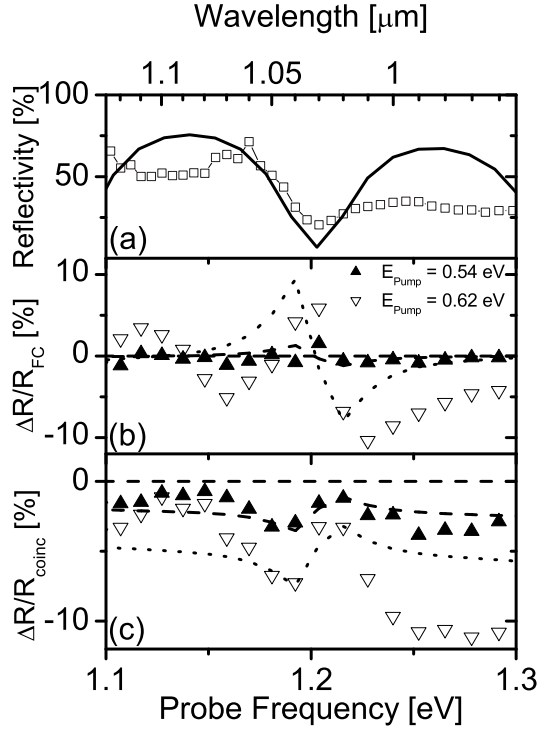}
\caption{\setstretch{2.0}(a): Measured linear reflectivity versus probe frequency (open squares) on domain D4 compared to the scalar wave approximation (solid
curve). (b) The differential reflectivity $(\Delta R/R)_{\rm{FC}}$ at delay $\Delta t = 1$ ps caused by the free carriers shows mostly
dispersive features, as seen from the symmetric variation of $\Delta R/R$ around 0. When increasing the pump frequency from 0.54 (solid
triangles) to 0.62 eV (open triangles), these dispersive features increase in magnitude. (c): At $\Delta t = 0$ ps, dispersive as well as
absorptive features are observed in $(\Delta R/R)_{\rm{coinc}}$, increasingly so when increasing the pump frequency from $E_{\rm{Pump}} = 0.54$
eV to $0.62$ eV. For $E_{\rm{Pump}} = 0.54$ eV, we find $\Delta n' = 0.44 \times 10^{-3}$ and $n'' = 0.46 \times 10^{-3}$, while for
$E_{\rm{Pump}} = 0.62$ eV the fits give $\Delta n' = 1.1 \times 10^{-3}$ and $n'' = 1.1 \times 10^{-2}$. The pump intensity was $I_{\rm{Pump}} =
36 \pm 4$ GWcm$^{-2} (E_{\rm{Pump}} = 0.54$ eV) and $I_{\rm{Pump}} = 46 \pm 5$ GWcm$^{-2} (E_{\rm{Pump}} = 0.62$ eV). Allt0psAllt1ps.eps}
\label{fig:dRRt0pst1ps}
\end{center}
\end{figure}

In order to investigate the precise nature of $(\Delta R/R)_{\rm{coinc}}$, we have measured the differential
reflectivity at coincidence  on domain D4 for a wide range of probe frequencies, which we plot in figure \ref{fig:dRRt0pst1ps}(c). We make three
important observations. Firstly, with increasing $E_{\rm{Pump}}$, the variation in the data becomes stronger. Secondly, the differential
reflectivity becomes increasingly negative with both increasing pump and increasing probe frequency. These negative values indicate optical
absorption. Thirdly, at 1.22 eV, a peak appears in the differential reflectivity that corresponds to a red edge in the linear reflectivity
(figure \ref{fig:dRRt0pst1ps}(a)); likewise, a trough appears at 1.18 eV which corresponds to a blue edge. This clear dispersive behavior is
indicative of a redshift of the photonic features related to a positive change in $n'$. Thus we conclude that when pump and probe are
coincident, two effects contribute to the differential reflectivity: Nonlinear dispersion, which we attribute to the electronic Kerr effect, and
absorption. We conclude that the negative differential reflectivity is due to an instantaneous nondegenerate two photon process, since this
absorption is seen to increase for an increase of either pump or probe frequencies, and since the sum of pump and probe frequencies exceeds the
absorption edge of p-Si. Recent work \cite{Hache:00, Mazurenko:03, Hastings:05, Mondia:05, KerrRetract:07} has attributed the behavior at
coincidence to the Kerr effect exclusively, even though dispersive data are lacking. Here, for the first time we identify the different
instantaneous contributions to all-optical photonic switching. Therefore, this also represents the first identification of Kerr switching in
photonic crystals.

\subsection{Reflectivity vs. frequency at $\Delta t = 1$ ps}
\label{subsec:Rvsomegat1} Before quantifying the two different instantaneous contributions from the data, we will first briefly describe the
measured effects of the free carriers on the spectral properties. Figure \ref{fig:dRRt0pst1ps}(b) shows the differential reflectivity at a
positive probe delay of $\Delta t = 1$ ps that is caused by free carriers. Because the differential reflectivity varies around 0, we conclude
that the induced absorption is minimal. The peak at 1.2 eV corresponds to the blue edge of the linear reflectivity peak at 1.175 eV, the two
troughs at 1.23 eV and 1.16 eV corresponds to the red edges of the reflectivity peaks at 1.175 and 1.25 eV, respectively. Therefore, the data
are consistent with a blueshift, or a decrease of the refractive index $\Delta n'$ as a result of the free carriers.

\subsection{Model: The Extended Scalar Wave Approximation}
\label{subsec:SWA} To quantify the two nonlinear contributions at coincidence as well as the free carrier effects, a model of the woodpile's
linear and nonlinear spectral properties is necessary. To predict the change in complex refractive index, we require a physical model which is
subject to two requirements: First, this model must reproduce the linear reflectivity $R_0$ over a given frequency range. This requirement is
necessary because the measured differential reflectivity, defined as $\frac{\Delta R(t)}{R} \equiv \frac{R(t) - R_0}{R_0}$, depends sensitively
on $R_0$. Second, the calculated spectral properties must scale correctly with the induced complex $n$.

Here, we employ a heuristic model to calculate both the linear as well as the dynamic reflectivity, which is an extension of the scalar wave approximation (SWA) \cite{Shung:93}. The dielectric function in beam direction $\epsilon(z)$ and the lattice spacing can be written
in terms of their Fourier components $U_G$ and $G$, respectively. For binary structures composed of two different dielectric materials, it can
be shown that \cite{KoenderinkThesis:03}
\begin{equation}
U_G = \Delta \epsilon f_G, G \neq 0, \label{eq:UG}
\end{equation}
where $\Delta \epsilon$ is the difference of the dielectric constants, and $f_G$ is the Fourier Transform of the indicator function $f(z)$,
where $\epsilon(z) = \epsilon_1 + \Delta \epsilon f(z)$. From a change in $\Delta \epsilon$, a new $U_G$ can readily be obtained. The electric
field inside the crystal can now be calculated by considering only the first two bands, $k$ (incident) and $k-G$ (Bragg diffracted). Outside the
crystal, the field can be determined from the two boundary conditions the two bands are subject to. Thus, the transmission and reflectivity can
be determined. Recently, Euser {\em{et al.}} employed an exact modal method (EMM) to calculate the broadband reflectivity of the woodpile
photonic crystal \cite{Gralak:03}. Although the agreement was good for low frequencies, the reflectivity could not be reproduced faithfully
beyond the stopgap \cite{EuserWoodpile:08}. To obtain a better agreement, the number of plane waves would have had to have been increased, for
which the calculation diverges slower. Also, Deubel {\em{et al.}} acheived good agreement at the blue edge with a scattering matrix model, although some salient features were still lacking \cite{Deubel:05}. We verified that at the relevant probe frequencies, the induced shift obtained with SWA matches the EMM
(see appendix \ref{app:EMMSWA}).

We fit the linear reflectivity obtained by the scalar wave approximation to the region given in fig. \ref{fig:dRRt0pst1ps}(a). The best
agreement to the measured reflectivity is obtained with $G=2\pi/(393$ nm) and $U_G = -0.8$. The calculated spectral feature shown corresponds to the first two Fabry-P\'erot
fringes at the blue edge of the stopgap. In the model, we included the dispersion and absorption of p-Si (from \cite{WVASE:32}, see fig.
\ref{fig:ExtractCmplxn}), and have taken bulk Si as the substrate. We verified that the thin layer of SiN ($n'=2, d=70$nm) had no effect on our
analysis (appendix \ref{app:SiN}). Since the optical features in the reflectivity spectrum are much broader than the probe bandwidth, no
convolution of the calculated spectrum with the probe is necessary.

\subsection{Interpretation of spectra at coincidence}
\label{subsec:InterpretationCoincidence} We match the calculated dispersive differential reflectivity to the strong signature of a redshift at
$E_{\rm{Probe}}=1.2$ eV by fitting two independent parameters $\Delta n' = 2 n_2 I_{\rm{Pump}}$ and $n'' =
\lambda_0\beta_{12}I_{\rm{Pump}}/2\pi$ over the range $1.18$ eV $< E_{\rm{Probe}} < 1.23$ eV by minimizing the least squares. Here, $\beta_{12}$
is the nondegenerate two-photon absorption coefficient and $\lambda_0$ the free space wavelength. The differential reflectivity was obtained as
follows: using the $G$ obtained previously from the fit to the linear spectrum, the $U_G$ was changed via the changed $\Delta \epsilon$
(equation \ref{eq:UG}), and the switched reflectivity spectrum was calculated from the new $U_G$. Error margins for the parameters were taken as
the values at which the least squares values had changed by $10 \%$.

Figure \ref{fig:dRRt0pst1ps}(c) shows the fits to two spectra obtained at two different $E_{\rm{Pump}}$. The remarkable agreement of this
approximate model with the data confirms that the subtle interplay of nonlinear dispersion as well as nondegenerate absorption leads to the
observed features. At frequencies below $1.2$ eV, the fitted curves tend to be lower than the measurements. Above $1.2$ eV, the calculated
differential reflectivity is mostly higher than the measurements. These deviations are due to our fit not taking into account the frequency
dependence of $\beta_{12}$, but setting it to one value for all probe frequencies.

\begin{figure}
\begin{center}
\includegraphics[scale=1.0]{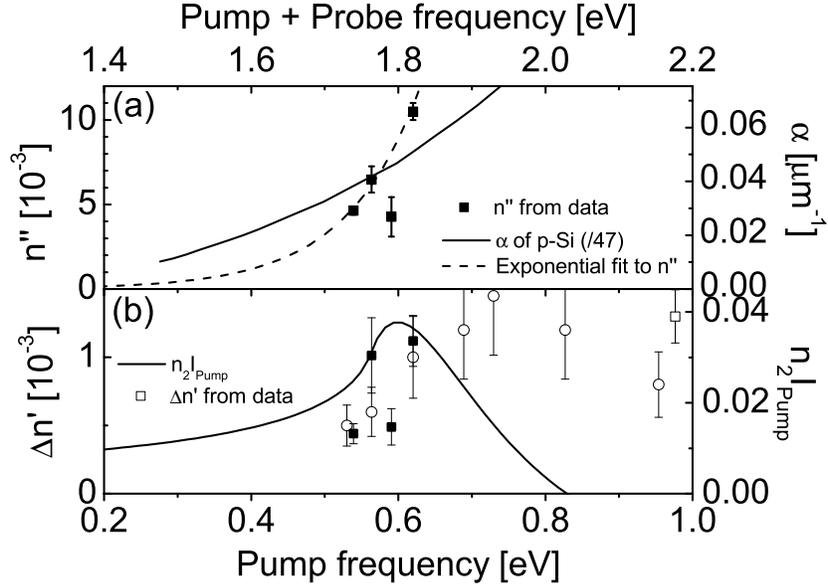}
\caption{\setstretch{2.0}Instantaneous changes of the complex refractive index. (a): Imaginary refractive index ($n''$, solid squares) vs. pump frequency
$E_{\rm{Pump}}$ (lower axis) obtained from fits in fig. \ref{fig:dRRt0pst1ps}. The dashed curve is a fit of $n''$ to an exponential. The right
scale shows the corresponding absorption coefficient. For comparison, we plot the absorption coefficient for low pressure chemical vapor
deposited p-Si (solid curve) annealed at $545^{\circ}$C, similar to the backbone of our woodpile crystals. Upper abscissa is the sum of the pump
and probe frequencies $E_{\rm{Total}}$. (b): Change in real refractive index $\Delta n'$ (left scale, solid squares) vs. pump frequency
$E_{\rm{Pump}}$ from our nondegenerate measurements. The theoretical relation predicted by \cite{SheikBahae:90} is also shown (solid curve,
right scale). We plot degenerate measurements for comparison: Open squares from \cite{Dinu:03} and open circles from \cite{Lin:07}. Data from
\cite{Bristow:07} exceeds the scale by one order of magnitude. ExtractCmplxn.eps}
 \label{fig:ExtractCmplxn}
 \end{center}
\end{figure}

Figure \ref{fig:ExtractCmplxn}(a) and (b) shows the absorptive and dispersive part of the induced refractive index at coincidence, respectively,
as returned from the fits. The value at $E_{\rm{Pump}} = 0.59$ eV is excluded due to an experimental artifact in that measurement. Figure
\ref{fig:ExtractCmplxn}(a) shows that the imaginary refractive index $n''$ increases for pump frequencies $E_{\rm{Pump}} > 0.55$ eV. These
frequencies correspond to a sum of pump and probe frequencies $E_{\rm{Pump}} + E_{\rm{Probe}} > 1.75$ eV. Since these values are larger than the
optical gap of p-Si ($1.4$ eV $< E_{\rm{opt}} < 1.6$ eV, depending on the degree of polycrystallinity \cite{Rotaru:99}), it is clear that the
simultaneous presence of pump and probe photons induces optical absorption. The resulting values of the nondegenerate two photon absorption
coefficient are $\beta_{12} = 0.11$ cmGW$^{-1}$ at $E_{\rm{Pump}} = 0.54$ eV and $\beta_{12} = 0.19$ cmGW$^{-1}$ at $E_{\rm{Pump}} = 0.62$ eV.

From recent calculations \cite{Garcia:06}, the $N_{\rm{ph}}$-photon process in a direct transition can be viewed as a single-photon process with
an energy gap rescaled to $E_G/N_{\rm{ph}}$, where $N_{\rm{ph}}$ is the number of photons. In other words, $(E_{\rm{Pump}} + E_{\rm{Probe}})/2 =
E'$ in the present two beam case, where $E'$ is a rescaled energy. Therefore, we can compare the spectral {\em{shape}} of our extracted
nondegenerate absorption coefficient to the linear absorption. We therefore also plot the linear absorption of p-Si \cite{WVASE:32}, similar to
our woodpile crystals \cite{Fleming:99}. The behavior of the linear absorption agrees reasonably with that of the nondegenerate absorption,
supporting calculations from \cite{Garcia:06}. For our intensities, the nondegenerate absorption remains significantly below the linear
absorption. Pumping or probing at higher frequencies causes the absorption to increase, as do higher pump intensities, but not higher probe
intensities \cite{ProbeIntensity}.

The instantaneous change of the real part of the refractive index $\Delta n'$ is plotted in figure \ref{fig:ExtractCmplxn}(b). Again the value
at $E_{\rm{Pump}} = 0.59$ eV is excluded from the discussion. At low pump frequencies of 0.52 eV, the change is $0.44 \times 10^{-3}$, and it
more than doubles to $1.1 \times 10^{-3}$ at $E_{\rm{Pump}} = 0.56$ eV. At $E_{\rm{Pump}} = 0.62$ eV, $\Delta n'$ decreases slightly. The
resulting values of the Kerr coefficients are $n_2 = 7.3 \times 10^{-6}$ cm$^2$GW$^{-1}$ at $E_{\rm{Pump}} = 0.54$ eV and $n_2 = 12.7 \times
10^{-6}$ cm$^2$GW$^{-1}$ at $E_{\rm{Pump}} = 0.62$ eV. We compare our measurements to recent measurements of the degenerate nonlinear dispersion
of bulk Si \cite{Dinu:03, Lin:07, Bristow:07}. In order to be able to make the comparison, we made use of the rescaling condition. We find
excellent agreement to the data from \cite{Dinu:03, Lin:07}, but surprisingly the measurements of \cite{Bristow:07} are an order of magnitude
higher, despite all being measured with the same z-scan technique. We also compare our measured data to a relation for $n_2$ derived for
direct-gap semiconductors \cite{SheikBahae:90} multiplied by our pump intensity $I_{\rm{Pump}}$. The functional form agrees well with our data.
At higher probe frequencies, the theoretical relation differs from the data of Refs. \cite{Lin:07, Dinu:03}. The agreement can be improved by
employing a theory for the degenerate nonlinear properties of an indirect semiconductor \cite{Garcia:06} using the rescaling condition
\cite{Bristow:07}.

Having identified the two contributions at pump-probe coincidence, we briefly discuss recent photonic crystal switching experiments in the light
of our new interpretation. Becker {\em{et al.}} performed time-resolved non-degenerate pump-probe spectroscopy on a-Si and p-Si inverse opals
\cite{BeckerAPL:05}. In their data on the a-Si sample, sharp features indicative of instantaneous absorption are apparent for $E_{\rm{Pump}} =
0.674$ eV and $E_{\rm{Probe}} > 0.653$ eV, or $E_{\rm{Total}} > 1.327$ eV at pump-probe coincidence. Their p-Si sample reveals these sharp
features at $E_{\rm{Pump}} = 0.717$ eV and $E_{\rm{Probe}} > 0.539$ eV ($E_{\rm{Total}} > 1.256$ eV). Because of the lower optical gap
$E_{\rm{Opt}}$ for p-Si with respect to a-Si, we expect nondegenerate absorption to take place at a lower $E_{\rm{Total}}$ than for the a-Si
sample, in agreement with their data. As a second example, we discuss the contribution by Hach\'e and Bourgeois \cite{Hache:00}. Here, the
authors probe a Si/SiO$_2$ Bragg stack at the red edge of the stopgap. At coincidence, they observe a significant increase in transmission. The
observed feature can only be explained by the breakdown of destructive interference due to nondegenerate absorption for $E_{\rm{Pump}} = 0.725$
eV and $E_{\rm{Probe}} = 0.821$ eV, giving $E_{\rm{Total}} = 1.546$ eV, comparable to $E_{\rm{Opt}}$ of a-Si. The Kerr effect without induced
absorption would lead to a red shift of the stopgap and thus to a decrease in transmission. In a similar manner, the contributions by Mazurenko
{\em{et al.}}, Hastings {\em{et al.}}, and Mondia {\em{et al.}} misinterpret the sharp trough in differential reflectivity as the Kerr effect
\cite{Mazurenko:03, Hastings:05, Mondia:05}.

\subsection{Interpretation of ps switching}
Using the same extended SWA from section \ref{subsec:SWA}, we can also interpret the picosecond free carrier behavior. Therefore, the calculated
differential reflectivity can be fitted to $(\Delta R/R)_{\rm{FC}}$ by calculating the complex refractive index change with the Drude model
(figure \ref{fig:dRRt0pst1ps}(b)). The carrier density $N$ is given by
\begin{equation}
N = \frac{I^2_{\rm{Pump}}\tau_P\beta_{11}}{2E_{\rm{Pump}}e}, \label{eq:density}
\end{equation}
where $\beta_{11}$ is the degenerate two photon absorption coefficient. At our pump frequencies, we can neglect linear absorption. For carrier
densities $N < 10^{28}$ m$^{-3}$, well above densities created in our experiments, the change in dielectric constant is
\cite{SokolowskiTinten:00}
\begin{equation}
\Delta \epsilon_{\rm{fc}} = -\left(\frac{\omega_P}{\omega}\right)^2\left(1 - i(\omega \tau_D)^{-1}\right),
 \label{eq:Drude}
\end{equation}
where $\omega_P^2 = Ne^2/(m^{*}\epsilon_0)$ is the plasma frequency squared in (rad/s)$^2$ and $e$ and $m^{*}$ are the charge and effective
optical mass of the carriers, and $\omega$ the probe frequency in rad/s. Equation \ref{eq:Drude} is valid for $(\omega\tau_D)^{-1}<<1$, which is
the case for our probe frequencies and damping times. For a given $I_{\rm{Pump}}$ therefore, we can fit the differential reflectivity resulting
from a changing $\beta_{11}$ to $(\Delta R/R)_{\rm{FC}}$.

\begin{figure}
\begin{center}
\includegraphics[scale=1.0]{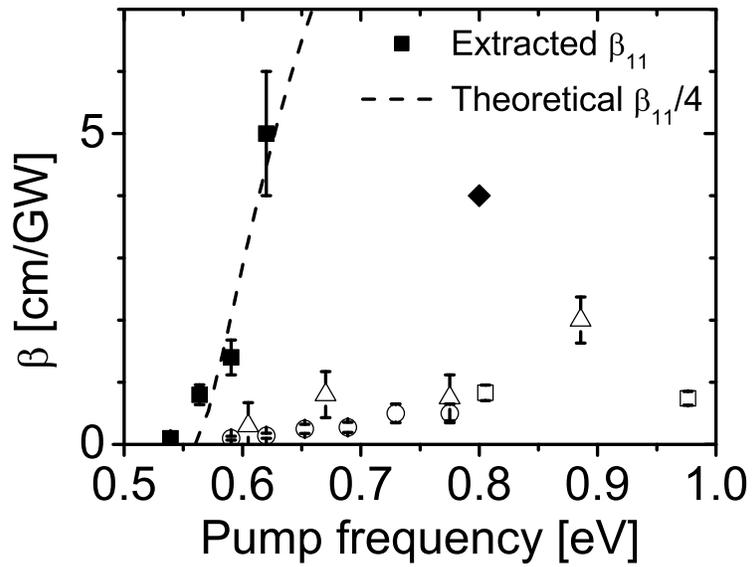}
\caption{\setstretch{2.0}Nonlinear coefficient $\beta_{11}$ vs. pump frequency (solid squares) obtained from data as in fig. \ref{fig:dRRt0pst1ps}. Dashed curve
is a calculation from Ref. \cite{SheikBahae:90} for c-Si. Other data are from \cite{Dinu:03} (c-Si, open squares), \cite{Lin:07} (c-Si, open circles),
\cite{Bristow:07} (c-Si, triangles) and \cite{Ikeda:07} (p-Si, diamonds). beta.eps}
 \label{fig:beta}
 \end{center}
\end{figure}

The $\beta_{11}$ versus $E_{\rm{Pump}}$ as returned from the fits is plotted in figure \ref{fig:beta}. For $E_{\rm{Pump}}<\frac{1}{2} E_G$,
$\beta_{11}$ tends to $0$, as expected. Increasing $E_{\rm{Pump}}$ to above $0.54$ eV, there is a marked increase. At $E_{\rm{Pump}} = 0.62$ eV,
$\beta_{11}$ has increased to as much as $5 \pm 1$ cmGW$^{-1}$. This value compares favorably to other values of p-Si samples: values as high as
$\beta_{11} = 60 \pm 15$ cmGW$^{-1}$ have been reported \cite{EuserThesis:07}. Indeed, a systematic study by Ikeda {\em{et al.}} showed the
nonlinear coefficient increases with increasing amorphousness of the sample, with $\beta_{11} = 4$ cmGW$^{-1}$ for p-Si samples, and up to 120
cmGW$^{-1}$ for amorphous silicon \cite{Ikeda:07}. To summarize this comparison, we plot our extracted values of $\beta_{11}$ along with
$\beta_{11}$ of both p-Si and c-Si sample in fig. \ref{fig:beta}.

We are now able to calculate crucial parameters to characterize the quality of the switching process in terms of absorption and homogeneity. We
find that the carrier induced absorption length is $\ell_{\rm{abs}} = 123 \pm 50~\mu$m, much longer than the Bragg length $L_B$, confirming that
the absorption is negligible indeed. The pump homogeneity length that describes the spatial homogeneity of carriers is $\ell_{\rm{hom}} =
(I_{\rm{Pump}}\beta_{11})^{-1} = 44 \pm 10~\mu$m, or $170 \pm 40~L_B$ at $E_{\rm{Pump}} = 0.62$ eV and $I_{\rm{Pump}} = 46 \pm 5$ GWcm$^{-2}$.
This calculation highlights the advantage of two-photon absorption \cite{Euser:05} to one-photon absorption in which absorption strongly limits
the pump homogeneity \cite{Mazurenko:03, Hastings:05, Fushman:07}. We conclude that for our switching conditions, both the pump and the probe
absorption length remain well above $L$, and that thus instantaneous switching yields favorable conditions for our pump and probe frequencies,
and pump intensities.

\section{Interpretation of switched reflectivity at one probe frequency}
\begin{figure}
\begin{center}
\includegraphics[scale=1]{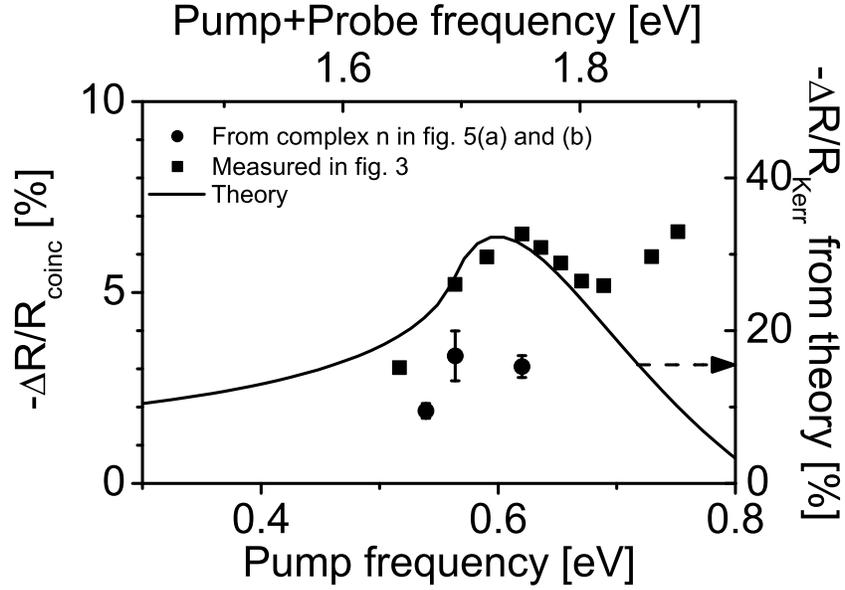}
\caption{\setstretch{2.0}Differential reflectivity at coincidence vs. pump frequency extracted from figure \ref{fig:timetraces} (domain A1) and corrected for
pump beam parameters (squares, left scale). Circles are complex values of $n$ extracted from fig. \ref{fig:ExtractCmplxn} (domain D4) and
reinserted in the extended SWA at domain A1. We have calculated the differential reflectivity expected from a nonlinear, purely dispersive
direct bandgap material (solid curve) \cite{SheikBahae:90} by inserting the theoretical relation for $n_2$ into the extended SWA. The 'purely
dispersive assumption' deviates from the data for pump frequencies above $E_{\rm{Pump}} = 0.69$ eV, or $E_{\rm{Total}} = 1.82$ eV.
KerrScaledwTheory.eps}
 \label{fig:dRR0pswithTheory}
 \end{center}
\end{figure}

Having analyzed spectra of the reflectivity at coincidence and at $\Delta t = 1$ ps, we return to figure \ref{fig:timetraces}. Experimental
conditions had prevented us from taking more than 4 probe frequency- and time-resolved spectra, mostly related to the long measurement times.
Having identified two competing processes at coincidence, we can now focus on time-resolved spectra which depend on pump frequency only. Figure
\ref{fig:dRR0pswithTheory} shows $(\Delta R/R)_{\rm{coinc}}$ as a function of $E_{\rm{Pump}}$. Here, $I_{\rm{Pump}}$ has been corrected for
three important pump beam parameters: i., the increase in intensity with increasing $E_{\rm{Pump}}$ because of the decreasing Gaussian focus
\cite{BeamDiam}; ii. the change in pump intensity due to the frequency dependent reflectivity of the pump, see figure \ref{fig:Refl}; and iii.
the measured $10 \%$ change in pump power over the frequency range of the OPA. For i., $(\Delta R/R)_{\rm{coinc}}$ is taken to vary
quadratically with $E_{\rm{Pump}}$, assuming that $(\Delta R/R)_{\rm{coinc}}$ changes linearly with $I_{\rm{Pump}}$. For ii, $(\Delta
R/R)_{\rm{coinc}}$ was linearly corrected by a factor $(1-R(E_{\rm{Pump}}))$, where $R_{\rm{Pump}}(E_{\rm{Pump}})$ is the reflectivity of the
pump at the pump frequency. Finally, in iii., we corrected $(\Delta R/R)_{\rm{coinc}}$ linearly with $I_{\rm{Pump}}$. The two largest
corrections are by $80 \%$ ($E_{\rm{Pump}} = 0.52$ eV) and $34 \%$ ($E_{\rm{Pump}} = 0.56$ eV), but otherwise the corrections are less than $20
\%$.

In figure \ref{fig:dRR0pswithTheory} we observe an increasing negative differential reflectivity $-(\Delta R/R)_{\rm{coinc}}$ with increasing
$E_{\rm{Pump}}$ up to $0.62$ eV, then a decrease until $0.69$ eV before a subsequent increase. With the analysis of the previous data (fig.
\ref{fig:ExtractCmplxn}), two important regimes can now be identified: for $E_{\rm{Pump}} < 0.69$ eV ($E_{\rm{Total}} < 1.82$ eV), $(\Delta
R/R)_{\rm{coinc}}$ is mostly governed by dispersion. For $E_{\rm{Pump}} > 0.69$ eV, $(\Delta R/R)_{\rm{coinc}}$ increases again due to a
combination of both dispersive ($n_2$) as well as absorptive parts ($\beta_{12}$).

In figure \ref{fig:dRR0pswithTheory} we also show the differential reflectivity expected from the extended SWA. Because this different
reflectivity spectrum was obtained on domain A1, we first have to apply the SWA to the linear reflectivity from A1 (not shown). We obtain best
agreement for $G=2\pi/(363$ nm) and $U_G = -1.2$. From the theoretical relation for a degenerate $n_2$ of a direct bandgap semiconductor
\cite{SheikBahae:90}, $(\Delta R/R)_{\rm{coinc}}$ is readily obtained. Since a nondegenerate theory for indirect bandgap semiconductors is not
yet available, the theory has to be interpreted cautiously with respect to the measurement. For $E_{\rm{Pump}}<0.7$ eV, the shape of the
theoretical relation agrees excellently to our measurements. A strong divergence from the data is found for $E_{\rm{Pump}}
> 0.7$ eV, or $E_{\rm{Total}} > 1.83$ eV. We interpret this threshold frequency as marking a transition from dispersive behavior of $(\Delta
R/R)_{\rm{coinc}}$ to absorptive behavior.

To verify the consistency of the time- and frequency resolved $(\Delta R/R)_{\rm{coinc}}$ taken on domain D4 (see section
\ref{subsec:Rvsomegat0}) with the pump frequency resolved $(\Delta R/R)_{\rm{coinc}}$ taken on domain A1, both $n_2$ and $\beta_{12}$ (figure
\ref{fig:ExtractCmplxn}) were inserted into the SWA at A1, and plotted in figure \ref{fig:dRR0pswithTheory}. The obtained differential reflectivity is lower by a factor $2$. Reasonable agreement to the
measured differential reflectivity is found, from which we conclude that the extended SWA provides consistent description of our various data,
measured at different pump intensities and on different sample domains.

\section{Nondegenerate figure of merit for fs switching}
\label{sec:NFOM} To quantify how useable a material is for optical switching, it is vital to consider a figure of merit (FOM). A high FOM then
is tantamount to a large phase shift, desirable for switching applications. Previously, Garmire proposed a degenerate FOM, a measure which
indicates the maximum achievable phase shift of $2 \pi \times$ FOM, given that the device length is equal to the pump homogeneity length
\cite{Garmire:94}. The absolute length of the device then depends on the intensity used. From this requirement it follows that,
\begin{equation}
{\rm{FOM}}(E_{\rm{Pump}}) = \frac{n_2(E_{\rm{Pump}})}{\lambda_{\rm{Pump}}\beta_{11}(E_{\rm{Pump}})},
\end{equation}
where $\lambda_{\rm{Pump}}$ is the pump wavelength. Here we extend this notion to the general case by including nondegenerate absorption: Not
only is the pump absorbed because it excites free carriers, but the probe is absorbed in the presence of the pump. The nondegenerate FOM (NFOM)
is then \cite{Lackn2}
\begin{equation}
\rm{NFOM}(E_{\rm{Pump}}, E_{\rm{Probe}}) =
\frac{n_2(E_{\rm{Pump}})}{\lambda_{\rm{Pump}}\beta_{11}(E_{\rm{Pump}}) + \lambda_{\rm{Probe}}\beta_{12}(E_{\rm{Pump}}, E_{\rm{Probe}})}.
\label{eq:NFOM}
\end{equation}
Because of the much lower probe intensity, we can neglect $n_2(E_{\rm{Probe}})$ in the numerator and $\beta_{11}(E_{\rm{Probe}})$ in the
denominator. The NFOM has been calculated from our measured data as follows: $n_2(E_{\rm{Pump}})$ has been derived from figure
\ref{fig:ExtractCmplxn}(b), where we have used the functional form of $n_2(E_{\rm{Pump}})$ from \cite{SheikBahae:90}, scaled by a constant
factor to match the magnitude of our data. An analytic form of $\beta_{12}$ was derived by fitting an exponential to the linear absorption
coefficient $\alpha(E_{\rm{Pump}} + E_{\rm{Probe}})$. The exponential behavior of the F absorption coefficient $\alpha$ with
$E_{\rm{Probe}}$ close to the gap region is well documented \cite{Klingshirn:05}. From $\alpha(E_{\rm{Pump}} + E_{\rm{Probe}})$, we deduce
$\beta_{12} = \alpha(E_{\rm{Pump}} + E_{\rm{Probe}})/(2I_{\rm{Pump}})$ from equation \ref{eq:diffeqnondegenabs}. Finally, an analytic expression
for $\beta_{11}(E_{\rm{Pump}})$ is obtained from the theoretical relation for $\beta_{11}(E_{\rm{Pump}})$ for direct bandgap semiconductors from
\cite{SheikBahae:90}, scaled by a factor to match our measurements (fig. \ref{fig:beta}).

\begin{figure}
\begin{center}
\includegraphics[scale=1]{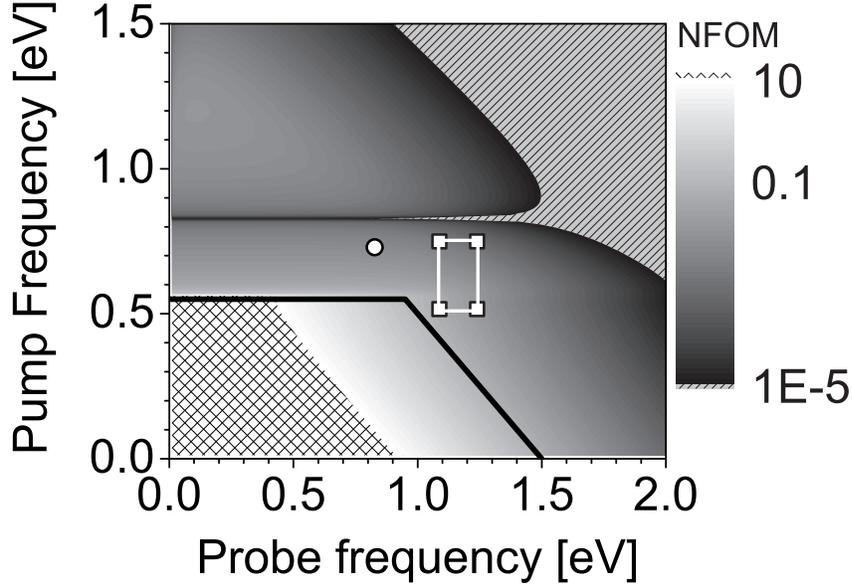}
\caption{\setstretch{2.0}Nondegenerate Figure of Merit (NFOM) for instantaneous switching vs. pump and probe frequency for polysilicon (see eq. \ref{eq:NFOM}).
For $E_{\rm{Pump}} < \frac{1}{2} E_G$ (=0.56 eV), pump absorption is absent as we do not consider more than second order photon absorption. If
in addition $E_{\rm{Pump}} + E_{\rm{Probe}} < E_{\rm{Opt}} = 1.5$ eV, the total absorption is 0 leading to large NFOM outside of the scale of
graph (crossed area). The black curve indicates the parameter space for which a maximum NFOM is reached if three photon absorption is neglected.
At $E_{\rm{Pump}} = 0.8$ eV, $n_2(E_{\rm{Pump}})$ crosses 0 cm$^{2}$GW$^{-1}$, and so the NFOM vanishes. For $E_{\rm{Total}} >> E_{\rm{Opt}}$,
the NFOM is $< 1 \times 10^{-5}$ (hatched area). The parameter space in our present measurements are bounded by the box (squares). The only
other measurement of $n_2$ on a Si based photonic structure is \cite{Hache:00} (circle). NFOM.eps} \label{fig:NFOM}
\end{center}
\end{figure}

Figure \ref{fig:NFOM} shows the NFOM for different $E_{\rm{Pump}}$ and $E_{\rm{Probe}}$. At total frequencies $E_{\rm{Total}} = E_{\rm{Pump}} +
E_{\rm{Probe}}$ below the optical gap of p-Si ($E_{\rm{Total}} < E_{\rm{Opt}} \approx 1.5$ eV), the nondegenerate two-photon absorption is
small, tantamount to a long probe absorption length, which is desired. If in addition $E_{\rm{Pump}}< \frac{1}{2} E_G$, the pump absorption
length becomes infinite. Higher order photon absorption has to be considered to quantitatively estimate the NFOM in this region of parameter
space \cite{Pearl:08}. Around $E_{\rm{Pump}} = 0.8$ eV, $n_2$ changes sign and thus the NFOM approaches $0$. For all other frequencies, NFOM
depends strongly on both pump and probe frequencies. To achieve a high NFOM, it is best for $E_{\rm{Total}}$ to remain below the optical gap,
while $E_{\rm{Pump}}$ should remain just below $\frac{1}{2} E_G$. In that case, $n_2$ is high, while both $\beta_{11}$ and $\beta_{12}$ are kept
low. We have indicated this line of maximum NFOM in fig. \ref{fig:NFOM}. These experimental conditions were recently realized by us on a
GaAs/AlAs microcavity \cite{Hartsuiker:08}. Our present measurements have taken place in the parameter space indicated by the box. In this
region, we find NFOM between $6 \times 10^{-4}$ for $E_{\rm{Probe}} = 1.24$ eV and $E_{\rm{Pump}} = 0.75$ eV, and NFOM $= 0.05$ for
$E_{\rm{Probe}} = 1.09$ eV and $E_{\rm{Probe}} = 0.58$ eV. The only other measurement of $n_2$ on a Si based photonic structure is from Ref.
\cite{Hache:00}. We conclude that for a judicious choice of pump and probe frequency, the NFOM can be maximized by limiting both pump and
nondegenerate probe absorption.

\section{Conclusion and recommendations}
Using nondegenerate tunable pump-probe spectroscopy on a Si woodpile photonic band gap crystal, we have for the first time identified Kerr
switching in photonic crystals. For pump$+$probe frequencies greater than the optical gap of p-Si, we additionally observe absorptive features,
which we identify with nondegenerate two photon absorption. Using a heuristic model, we estimate the maximum change in the real part of the
refractive index to be $\Delta n' = 1.1 \times 10^{-3}$ , giving a Kerr coefficient of $n_2 = 17 \times 10^{-6}$, which is in good agreement to
values in literature. The maximum imaginary part is $1.1 \times 10^{-2}$, corresponding to $\beta_{12} = 1.19$ cmGW$^{-1}$ at $E_{\rm{Pump}} =
0.62$ eV. This complex refractive index gives rise to reflectivity changes of $1 \%$. From the interplay of these two effects we derive a
nondegenerate figure of merit for instantaneous switching (NFOM). For a total pump+probe frequency in excess of the optical gap of Si, the NFOM
is low, while opportune switching is achieved for low probe frequencies, and for $E_{\rm{Pump}} \sim 1/2 E_G$.

At probe delays of $\Delta t = 1$ ps, we find that both pump and probe absorption is negligible, in agreement with a Drude description of the
free carriers. The differential reflectivity spectra is purely dispersionlike, giving changes in reflectivity of up to $\Delta R = 2.5 \%$. From
this value we deduce the maximum two-photon absorption coefficient to be $\beta_{11} = 5$ cmGW$^{-1}$, in agreement to values stated in
literature for p-Si.

In spite of the challenges in analyzing nondegenerate optical properties of a photonic band gap crystal, we have measured a clear difference in
response of $E_{\rm{Total}} < E_{\rm{Opt}}$ and $E_{\rm{Total}} > E_{\rm{Opt}}$. This difference has ramifications on future experiments on
nondegenerate instantaneous switching photonic bandgap crystals. DOS switching, and ultimately spontaneous emission switching is limited by the
magnitude of pump and emission frequencies.

To exploit Kerr switching, that is to maximize NFOM, following improvements are proposed. In first instance it seems useful to pump and probe
with the same polarization. For collinear polarization, a factor 3 improvement is predicted with respect to orthogonal polarization
\cite{SheikBahae:94}. Furthermore, the optical properties of the photonic crystal could be tailored so as to enhance the field. Calculations by
us on a 1D Distributed Bragg Reflector have revealed that the field can significantly be enhanced when pumping at either the red or blue
stopband edge, while the position of the field maximum depends critically on the exact pump frequency.

While a high-bandgap semiconductor will be less prone to nondegenerate absorption, $n_2$ scales with $E_G^{-3}$, thus reducing the NFOM. The
absorption can also be diminished by a lower pump intensity, which also goes at the expense of $\Delta n'$. We therefore
conclude that judicious choice of semiconductor material, and pump and probe frequencies are required to usefully exploit Kerr switching.

\section{Acknowledgements}
We thank Henry van Driel and Allard Mosk for discussions, and Albert Polman for making the samples available. This work is part of the research
program of the "Stichting voor Fundamenteel Onderzoek der Materie (FOM)", which was supported by the "Nederlandse Organisatie voor
Wetenschappelijk Onderzoek (NWO)".

\bibliographystyle{osajnl}

\begin{thebibliography}{10}
\bibitem{KerrRetract:07}
Indeed, this Kerr nonlinearity was also claimed by our group
  \cite{EuserArxiv:06}, but was later corrected \cite{EuserWoodpile:08}.
\bibitem{BeamDiam}
We had independently measured the radius of the beam waist at the focus, and
  had confirmed that the radius is diffraction limited.
\bibitem{Lackn2}
We note that the $n_2$ in equation \ref{eq:NFOM} depends on $E_{\rm{Pump}}$
  only. Measurements of nondegenerate $n_2$ are still lacking in literature.
\end{thebibliography}

\clearpage

\section*{List of Figure Captions}
Fig. 1 High resolution scanning electron micrograph of the surface normal to [001] of a Si woodpile crystal at domain D4. The width and
thickness of each rod is $175 \pm 10$nm and $155 \pm 10$nm, respectively.

Fig. 2 Linear reflectivity spectrum of the woodpile photonic crystal measured normal to the [001] direction at a sample domain D4 shown in
figure \ref{fig:SEM}. The E-field is perpendicular to the first row of rods. A stopgap near 0.9 eV gives rise to a high maximum reflectivity of
$95 \pm 2 \%$ and has a broad relative width of 47$ \%$, indicating a high photonic strength. At high frequencies $> 1.2$ eV, the spectral
features are attributed to Fabry-P\'{e}rot-type fringes. The pump frequencies (shaded box) were tuned at the red edge through half the
electronic band gap $E_{\rm{G}}=1.12$ eV of silicon (vertical dashed line), and the probe frequencies at the blue edge of the stopband. The
dashed curve is a calculation with the Scalar Wave Approximation in the region of interest.

Fig. 3 Differential reflectivity $\Delta R/R$ versus probe delay $\Delta t$ taken at different pump frequencies $E_{\rm{Pump}}$ and at probe
frequency $E_{\rm{Probe}} = 1.13$ eV, sample domain A1. At $\Delta t = 0$ ps, the pump and probe are coincident, and the differential
reflectivity $(\Delta R/R)_{\rm{coinc}}$ decreases. At $\Delta t=1$ ps, the differential reflectivity $(\Delta R/R)_{\rm{FC}}$ has increased due
to the dispersion of the free carriers (FC). The peak pump intensity varies between $I_{\rm{Pump}} = 10 \pm 1$ GWcm$^{-2} (E_{\rm{Pump}} =
0.516$ eV) and $I_{\rm{Pump}} = 25 \pm 2$ GWcm$^{-2} (E_{\rm{Pump}} = 0.75$ eV), and the probe intensity was $I_{\rm{Probe}} = 3\pm 2$
GWcm$^{-2}$.

Fig. 4 (a): Measured linear reflectivity versus probe frequency (open squares) on domain D4 compared to the scalar wave approximation (solid
curve). (b) The differential reflectivity $(\Delta R/R)_{\rm{FC}}$ at delay $\Delta t = 1$ ps caused by the free carriers shows mostly
dispersive features, as seen from the symmetric variation of $\Delta R/R$ around 0. When increasing the pump frequency from 0.54 (solid
triangles) to 0.62 eV (open triangles), these dispersive features increase in magnitude. (c): At $\Delta t = 0$ ps, dispersive as well as
absorptive features are observed in $(\Delta R/R)_{\rm{coinc}}$, increasingly so when increasing the pump frequency from $E_{\rm{Pump}} = 0.54$
eV to $0.62$ eV. For $E_{\rm{Pump}} = 0.54$ eV, we find $\Delta n' = 0.44 \times 10^{-3}$ and $n'' = 0.46 \times 10^{-3}$, while for
$E_{\rm{Pump}} = 0.62$ eV the fits give $\Delta n' = 1.1 \times 10^{-3}$ and $n'' = 1.1 \times 10^{-2}$. The pump intensity was $I_{\rm{Pump}} =
36 \pm 4$ GWcm$^{-2} (E_{\rm{Pump}} = 0.54$ eV) and $I_{\rm{Pump}} = 46 \pm 5$ GWcm$^{-2} (E_{\rm{Pump}} = 0.62$ eV).

Fig. 5 Instantaneous changes of the complex refractive index. (a): Imaginary refractive index ($n''$, solid squares) vs. pump frequency
$E_{\rm{Pump}}$ (lower axis) obtained from fits in fig. \ref{fig:dRRt0pst1ps}. The dashed curve is a fit of $n''$ to an exponential. The right
scale shows the corresponding absorption coefficient. For comparison, we plot the absorption coefficient for low pressure chemical vapor
deposited p-Si (solid curve) annealed at $545^{\circ}$C, similar to the backbone of our woodpile crystals. Upper abscissa is the sum of the pump
and probe frequencies $E_{\rm{Total}}$. (b): Change in real refractive index $\Delta n'$ (left scale, solid squares) vs. pump frequency
$E_{\rm{Pump}}$ from our nondegenerate measurements. The theoretical relation predicted by \cite{SheikBahae:90} is also shown (solid curve,
right scale). We plot degenerate measurements for comparison: Open squares from \cite{Dinu:03} and open circles from \cite{Lin:07}. Data from
\cite{Bristow:07} exceeds the scale by one order of magnitude.

Fig. 6 Nonlinear coefficient $\beta_{11}$ vs. pump frequency (solid squares) obtained from data as in fig. \ref{fig:dRRt0pst1ps}. Dashed curve
is a calculation from Ref. \cite{SheikBahae:90} for c-Si. Other data are from \cite{Dinu:03} (c-Si, open squares), \cite{Lin:07} (c-Si, open
circles), \cite{Bristow:07} (c-Si, triangles) and \cite{Ikeda:07} (p-Si, diamonds).

Fig. 7 Differential reflectivity at coincidence vs. pump frequency extracted from figure \ref{fig:timetraces} (domain A1) and corrected for pump
beam parameters (squares, left scale). Circles are complex values of $n$ extracted from fig. \ref{fig:ExtractCmplxn} (domain D4) and reinserted
in the extended SWA at domain A1. We have calculated the differential reflectivity expected from a nonlinear, purely dispersive direct bandgap
material (solid curve) \cite{SheikBahae:90} by inserting the theoretical relation for $n_2$ into the extended SWA. The 'purely dispersive
assumption' deviates from the data for pump frequencies above $E_{\rm{Pump}} = 0.69$ eV, or $E_{\rm{Total}} = 1.82$ eV.

Fig. 8 Nondegenerate Figure of Merit (NFOM) for instantaneous switching vs. pump and probe frequency for polysilicon (see eq. \ref{eq:NFOM}).
For $E_{\rm{Pump}} < \frac{1}{2} E_G$ (=0.56 eV), pump absorption is absent as we do not consider more than second order photon absorption. If
in addition $E_{\rm{Pump}} + E_{\rm{Probe}} < E_{\rm{Opt}} = 1.5$ eV, the total absorption is 0 leading to large NFOM outside of the scale of
graph (crossed area). The black curve indicates the parameter space for which a maximum NFOM is reached if three photon absorption is neglected.
At $E_{\rm{Pump}} = 0.8$ eV, $n_2(E_{\rm{Pump}})$ crosses 0 cm$^{2}$GW$^{-1}$, and so the NFOM vanishes. For $E_{\rm{Total}} >> E_{\rm{Opt}}$,
the NFOM is $< 1 \times 10^{-5}$ (hatched area). The parameter space in our present measurements are bounded by the box (squares). The only
other measurement of $n_2$ on a Si based photonic structure is \cite{Hache:00} (circle).

Fig. 9 (a) Relative change in frequency $\Delta E/E_0$ for a given change in the relative average refractive index $\Delta
n_{\rm{av}}/n_{\rm{av}}$. For comparison, the change according to Bragg's law $\Delta E/E_0 = -\Delta n_{\rm{av}}/n_{\rm{av}}$. (b) Change in
reflectivity for a given change in imaginary part of the refractive index $n''$. C

Fig. 10 Reflectivity calculated with the scalar wave approximation, with and without the thin layer of SiN.

\clearpage

\appendix
\section{Comparison of the scalar wave approximation to the exact modal method}
\label{app:EMMSWA} 
\begin{figure}
\begin{center}
\includegraphics[scale=0.7]{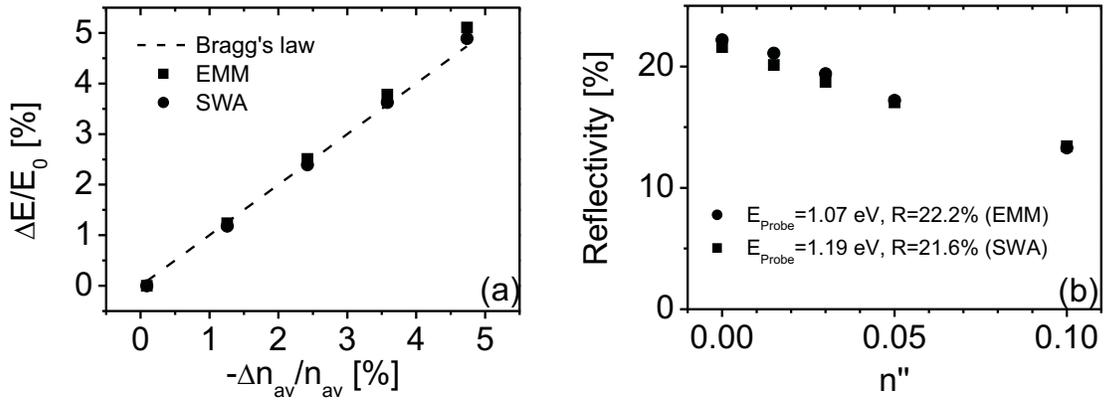}
\caption{\setstretch{2.0}(a) Relative change in frequency $\Delta E/E_0$ for a given change in the relative average refractive index $\Delta
n_{\rm{av}}/n_{\rm{av}}$. For comparison, the change according to Bragg's law $\Delta E/E_0 = -\Delta n_{\rm{av}}/n_{\rm{av}}$. (b) Change in
reflectivity for a given change in imaginary part of the refractive index $n''$. CmpEMMSWA.eps}
\label{fig:CmpEMMSWA}
\end{center}
\end{figure}

To assess whether the changes in spectral properties resulting from a change in $n'$ and $n''$ agree with the induced changes
in the EMM, we compare the spectral changes for a given $n'$ or $n''$. Figure \ref{fig:CmpEMMSWA}(a) shows the relative change in frequency for a
given change in average refractive index. The shift $\Delta E$ due to $\Delta n'$ at the same probe frequency is compared. We find that for
changes $\Delta n' < 0.2$, which is more than two orders of magnitude larger than our expected $\Delta n' = 1 \times 10^{-3}$, the agreement is
even better than $5 \%$. The good agreement for both $\Delta n'$ and $\Delta n''$ proves indeed that the two requirements are sufficient for the
description of the dynamic reflectivity. Therefore, spectral properties associated with the vectorial nature of the probe, such as TE-TM mixing
\cite{Paddon:00} or polarization effects \cite{deDood:03}, have a negligible effect on the {\em{change}} in reflectivity. In figure
\ref{fig:CmpEMMSWA}(b) we show the expected change in reflectivity for a given change in $n''$. To include absorption in the SWA, the
reflectivity is reduced by $\exp(-2L/\ell_{\rm{abs}})$, where $\ell_{\rm{abs}} = \lambda_0/(4\pi n'' \phi)$. The factor of $2$ in the exponent
accounts for the beam passing through the sample twice. For $E_{\rm{Probe}} = 1.19$ eV, and for $\Delta n'' < 0.1$, which is around 10 times
larger than expected, the agreement is better than $5 \%$.

From the good agreement between the exact modal method (EMM) and the extended scalar wave approximation, we conclude that the heuristic
accurately predicts the change in complex refractive index.

\section{Effect of SiN on linear reflectivity spectrum}
\label{app:SiN} 
\begin{figure}
\begin{center}
\includegraphics[scale=0.7]{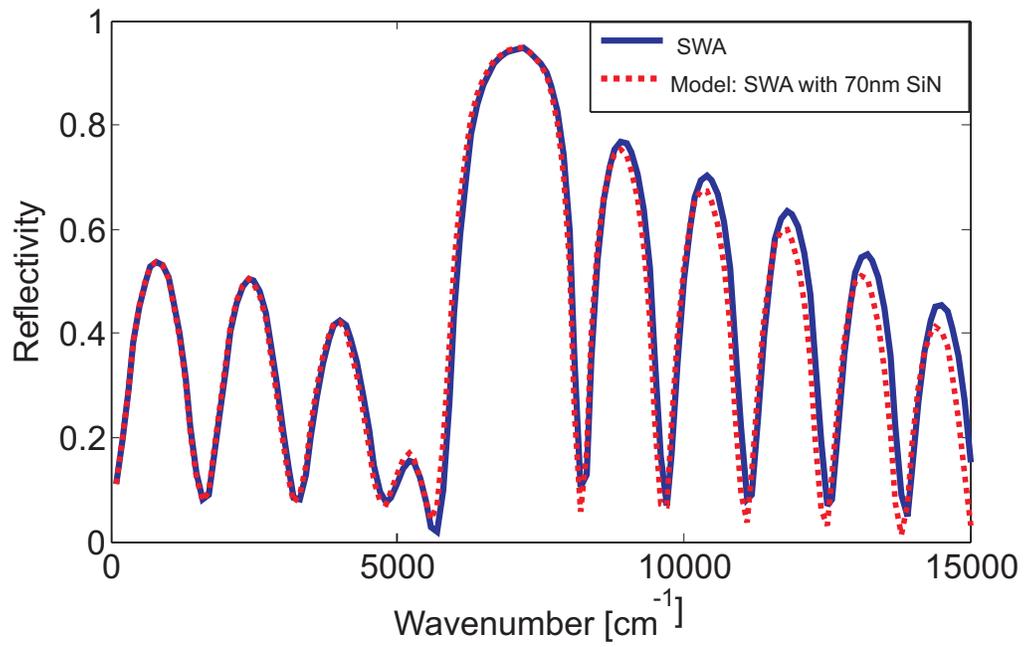}
\caption{\setstretch{2.0}Reflectivity calculated with the scalar wave approximation, with and without the thin layer of SiN. CmpSiNnoSiN.eps}
\label{fig:CmpSiNnoSiN}
\end{center}
\end{figure}

In order to verify that the layer of SiN supporting the woodpile has negligible effect on the linear reflectivity, a calculation
of the SWA is performed with this layer. Figure \ref{fig:CmpSiNnoSiN} shows the reflectivity spectra, without and with the supporting SiN layer
(70 nm). The spectra shifts by 6.2 meV at the most in our region of interest. Therefore we conclude that the inclusion of this layer has only a
negligible effect on the linear spectra.

\section*{Figures}

\end{document}